\documentclass[12pt,a4paper]{iopart}
\usepackage{iopams}
\usepackage[english]{babel}
\usepackage{float}
\usepackage[dvips]{graphicx}
\usepackage{rotating}
\usepackage{epsfig}
\usepackage[dvips]{color}
\definecolor{Red}{named}{Red}
\begin{document}
\title[Resolving Cosmic Neutrino Structure]{Resolving Cosmic Neutrino Structure: A Hybrid Neutrino $N$-body Scheme}
\author{Jacob Brandbyge$^1$, Steen Hannestad$^1$}
\ead{jacobb@phys.au.dk, sth@phys.au.dk}
\address{$^1$Department of Physics and Astronomy, University
of Aarhus, Ny Munkegade, DK-8000 Aarhus C, Denmark}
\date{\today}

%%%%%%%%%%%%%%%%%%%%%%%%%%%%%%%%%%%%%%%%%%%%%%%%%%%%%%%%%%%%%%%%%%%%%%%%%%%%%%%%%%%%%%%%%%%%%%%%%%%%%%%%%%%%%%%%%%%%%%%%%%%%%%%%%%%%%
\begin{abstract}
We present the first simulation capable of resolving the structure of neutrino clustering on Mpc scales. The method combines grid- and particle-based methods and achieves very good accuracy on both small and large scales, while keeping CPU consumption under control. Such simulations are not only ideal for calculating the non-linear matter power spectrum but also particularly relevant for studies of how neutrinos cluster in galaxy- or cluster-sized halos. We perform the largest neutrino $N$-body simulation to date, effectively containing 10 different neutrino hot dark matter components with different thermal properties.
\end{abstract}
\maketitle

%%%%%%%%%%%%%%%%%%%%%%%%%%%%%%%%%%%%%%%%%%%%%%%%%%%%%%%%%%%%%%%%%%%%%%%%%%%%%%%%%%%%%%%%%%%%%%%%%%%%%%%%%%%%%%%%%%%%%%%%%%%%%%%%%%%%%%
\section{Introduction}
%%%%%%%%%%%%%%%%%%%%%%%%%%%%%%%%%%%%%%%%%%%%%%%%%%%%%%%%%%%%%%%%%%%%%%%%%%%%%%%%%%%%%%%%%%%%%%%%%%%%%%%%%%%%%%%%%%%%%%%%%%%%%%%%%%%%%%

Neutrinos are among the most abundant particles in our Universe, and since at least two of the three neutrino mass eigenstates have masses much larger than the current temperature, neutrinos contribute to the matter density and are important for cosmological structure formation. Oscillation experiments have established two mass differences,
$\Delta{m}^2_{12}\simeq 7.6 \times
10^{-5} \, {\rm eV}^2$ and $|\Delta{m}^2_{23}| \simeq 2.4 \times 10^{-3} \, {\rm eV}^2$ \cite{Schwetz:2008er}, from which the heaviest mass eigenstate must have a mass of at least $0.05$ eV.
Even such a relatively small mass will have the effect of suppressing the power spectrum of matter fluctuations by $\sim 5$\%, an effect which is significantly larger than the precision with which the matter power spectrum can be measured in upcoming surveys.

Consequently this has led to a much increased interest in gaining a detailed understanding of how neutrinos affect structure formation. In linear theory the effect is extremely well understood. However, these results apply only on very large scales, $k \ll 0.1 \, h {\rm Mpc}^{-1}$, whereas most of the cosmologically relevant information from large-scale structure surveys is in the range $k \sim 0.1-0.5 \, h {\rm Mpc}^{-1}$.
On these intermediate scales it is mandatory to correct for non-linear effects, even if surveys are carried out at intermediate redshifts where non-linearity is weaker. In the semi-linear regime the power spectrum can be estimated using analytic methods (see \cite{Wong:2008ws,Lesgourgues:2009am,Saito:2009ah}).

However, in the fully non-linear regime the most accurate, but also most time-consuming method is to use full $N$-body simulations with neutrinos included in a self-consistent way. In a previous publication we carried out a large suite of simulations with neutrinos included \cite{Brandbyge1}, and found a significant non-linear correction caused by neutrinos.

In a recent paper \cite{Brandbyge2} we showed that for realistic neutrino masses, quantities such as the power spectrum can be calculated very reliably on all scales, using a grid-based method which tracks the linear neutrino density contrast on a grid while using the full non-linear structure of CDM.

The problem with both the particle-based and the grid-based methods is that they cannot reliably probe the small-scale structure of neutrinos. In particle-based codes the problem is that the thermal velocity of the neutrino component is so large that it introduces noise at an unacceptable level.
The grid-based method does not resolve neutrino bound structures and therefore by construction does not allow us to probe for example the neutrino content of a halo.

In the present paper we present a hybrid method which retains the good features of both methods and still runs at an acceptable speed. The idea is to start with neutrinos described on a grid, and then convert part of the grid to $N$-body particles when the thermal motion of neutrinos decreases to a few times the flow velocities in the simulation. The neutrino $N$-body particles are created in different momentum bins, with individual transfer functions and thermal velocities. We therefore present $N$-body simulations with 15 different neutrino hot dark matter species. In a following paper we will use this hybrid method to investigate neutrino clustering on sub Mpc scales.

In section~\ref{theory} we outline the theoretical framework and problems with combining particle and fluid approaches. Section~\ref{implementation} gives a technical description of the implementation of the hybrid method and Section~\ref{results} presents our results. Finally, Section~\ref{conclusions} contains our discussion and conclusions.

\section{Theory}\label{theory}
\subsection{The Boltzmann equation}
In this subsection we will briefly outline how the evolution of massive neutrinos is followed in linear theory. The notation is identical to \cite{Ma}. We use the metric in the conformal Newtonian gauge

\begin{equation}
  ds^2 = -a^2(1+2\psi)d\tau^2 + a^2(1-2\phi)dx^2.
  \label{eq:metric}
\end{equation}

In a perturbed universe the phase-space distribution function is expanded to first order as follows
\begin{equation}
  f= f_0 + \frac{\partial f_0}{\partial T} \delta T = f_0(1+\Psi),
  \label{eq:phase_space}
\end{equation}
with the perturbation parametrised by $\Psi = - d {\rm ln} f_0 / d {\rm ln} q ~ \delta T / T$, and the zeroth-order Fermi-Dirac phase-space distribution function given by
\begin{equation}
  f_0(q) = \frac{1}{e^{q/T} + 1}.
  \label{eq:f0}
\end{equation}
Here $q_i = a p_i$, where $p_i$ is the proper redshifting momentum and $T$ is the neutrino temperature today.

After neutrino decoupling the collisionless Boltzmann equation evolves the distribution function as

\begin{equation}
  \frac{df}{d\tau} = \frac{\partial f}{\partial \tau} + \frac{dx^i}{d\tau}\frac{\partial f}{\partial x^i}
   + \frac{dq}{d\tau}\frac{\partial f}{\partial q} + \frac{dn_i}{d\tau}\frac{\partial f}{\partial n_i} = 0.
  \label{eq:Boltzmann}
\end{equation}

Expanding the perturbation $\Psi$ in a Legendre series the perturbed neutrino energy density is given by a weighed sum over neutrino momentum states

\begin{equation}
  \label{eq:drho}
  \delta \rho _ \nu (k) = 4\pi a^{-4} \int{q^2dq\epsilon f_0 \Psi_0},
\end{equation}
where $\epsilon = (q^2 + a^2 m^2)^{1/2}$. Note that $f_0$ and not $f$ is used to weigh the individual momentum bins, an assumption which breaks down when the gravitational flow velocity approaches the thermal velocity of the individual momentum states.

From the Boltzmann equation the $\Psi_l$'s are related to each other and the metric potentials by

\begin{eqnarray}
\label{eq:Psi_l0}
\dot\Psi_0 & = & -\frac{qk}{3\epsilon}\Psi_1 - \dot\phi \frac{d{\rm ln}f_0}{d{\rm ln}q}, \\
\dot\Psi_1 & = & \frac{qk}{\epsilon} \biggl(\Psi_0 - \frac{2}{5}\Psi_2 \biggr) - \frac{\epsilon k}{q}\psi \frac{d{\rm ln}f_0}{d{\rm ln}q},\\
\dot\Psi_l & = & \frac{qk}{\epsilon} \biggl(\frac{l}{2l-1}\Psi_{l-1} - \frac{l+1}{2l+3}\Psi_{l+1} \biggr),~l\ge 2.
\label{eq:Psi_l2}
\end{eqnarray}
The second term on the right hand side of the equation for $\dot\Psi_0$ encodes the effect of structure forming in evolving gravitational potentials, whereas the first term incorporates the effect of velocity on structure formation. The change in velocity is affected by the three terms on the right hand side of the equation for $\dot\Psi_1$. The first term encodes the effect of flow velocities redshifting in an expanding Universe, whereas the second term, found from the hierarchy of $\dot\Psi_l$'s, incorporates the effect of momentum (and redshift) dependent neutrino free-streaming. These two terms give rise to less structure, whereas the last term in the equation for $\dot\Psi_1$ gives the acceleration as a gradient of the gravitational potential.

The quantity calculated in linear theory is the transfer function (TF) which evolves the initial primordial perturbation $\Psi_0^I$ as
\begin{equation}
  \Psi_0(k,q,z) = T(k,q,z) \Psi_0^I(k,q),
  \label{eq:TF}
\end{equation}
with $\Psi_0^I$ given by
\begin{equation}
  \Psi_0^I = - \frac{\delta T}{T} \frac{d {\rm ln} f_0}{d {\rm ln} q}.
  \label{eq:ic}
\end{equation}
With adiabatic initial conditions $\delta T / T = \delta_\nu / 4$ and $\delta_{\rm cdm} = \delta_{\rm b} = \frac{3}{4}\delta_\nu = \frac{3}{4}\delta_\gamma$ \cite{Ma}.

\subsection{Qualitative behaviour}
Eq.~(\ref{eq:ic}) states that neutrinos with different initial momenta have different initial perturbations, also on large scales. Specifically, $\Psi_0^I \rightarrow 0$ for $q / T \rightarrow 0$, meaning that neutrinos with zero (read: almost zero) momentum are unperturbed. The reason for this is that at each spatial location the neutrino distribution is initially locally a Fermi-Dirac distribution with zero chemical potential, as is also reflected in $\Psi_0^I$.
As $q/T \to 0$ the Fermi-Dirac distribution always tends to $f \to 1/2$, independent of $T$, and this means that the distribution function is unperturbed.

According to the Boltzmann equation, neutrinos with $q=0$ always stay unperturbed. When gravity acts on neutrino particles with $q=0$, thereby perturbing them, their momenta change, and they are therefore shifted out of the $q=0$ momentum bin, leaving this bin unperturbed. This fact is contained in the Boltzmann hierarchy for the $\Psi_l$'s through the term $d{\rm ln}f_0 / d{\rm ln} q$, which multiplies the gravitational potentials ($\phi$, $\psi$) so that different neutrino momentum bins are affected differently by gravity \footnote{Note that this is purely a phase-space phenomenon and has nothing to do with the gravitational force felt by particles.}.

The effect of linear as well as non-linear forces on the momenta of neutrino particles changes the shape of $f_0$ beyond linear order. Therefore the weighing of the individual $\Psi_0$'s in the momentum integral changes. This means that the value of $\delta \rho$ changes, not only at non-linear wavenumbers but also in the linear regime. Therefore the linear theory average neutrino TF is not even valid at very large scales, though becoming more accurate for lower neutrino masses.

\subsection{Converting the density grid to $N$-body particles}

When we convert densities on the grid into particles we are faced with a potential problem. Imagine that neutrino energy in the first $i$ momentum bins are converted to particles at some redshift, while keeping the remaining neutrinos on the grid. At some later time, some of the created particles will have been moved to higher velocities by gravity and therefore really belong to the grid, whereas some density on the grid would have moved to lower velocity, and should have been converted to particles. This ``leakage'' is problematic because the evolution of the grid is done in CAMB \cite{CAMB} while the evolution of particles is followed in \textsc{gadget}-2 \cite{Springel2}.

The problem is even more severe because the leakage to higher and lower momentum bins are qualitatively different: Imagine we dump all neutrino grid bins with momenta less than $q_{\rm cut}$ to $N$-body particles at a certain redshift, and let the particles evolve. Due to energy conservation, particles gaining energy above $q_{\rm cut}$ are falling into gravitational potential wells, and are typically clustered, while particles loosing energy are moving out of potential wells, and they are typically much more homogeneous as a population. Hence one cannot simply artificially decrease the velocities of the high energy particles to account for the leakage across $q_{\rm cut}$ between grid and particles, because the topology, or specific position in phase-space, of particles crossing the momentum boundary from above and below is very different.

Ideally, the neutrino $N$-body particles should only sample momentum space up to $q_{\rm cut}$. In the $N$-body simulation neutrinos with momenta larger than $q_{\rm cut}$ should therefore not contribute to the gravitational source term (to avoid double counting). This can surely be done, by eg. zeroing the mass of these neutrino $N$-body particles. Furthermore, we get a zero counting of the particles which, if they had been created from the grid at $q > q_{\rm cut}$, would have moved to $q < q_{\rm cut}$. One could include zero mass ghost particles with initial momentum states covered by the grid as tracer particles in the $N$-body volume, and then turn their mass on if they leak down to momentum states smaller than $q_{\rm cut}$. But in practise the linear assumption of a constant $f\simeq f_0$ is violated beyond linear order in the $N$-body simulation: There is a significant net flow of neutrino mass to higher momentum states. Therefore, turning the neutrino mass on / off as particles cross $q_{\rm cut}$ will lead to a net reduction of overall neutrino mass. Furthermore, particles moving to $q > q_{\rm cut}$ carry large non-linear corrections which are not found in the corresponding linear grid bins, making the idea with ghost particles less appealing.

However, a simple solution which works well in practise is to convert a sufficiently large number of momentum bins at the same time because that keeps the leakage between converted and non-converted bins to a minimum. We will discuss this issue in the next sections and demonstrate that errors can be kept under control at the required precision.

\section{Implementation of the Hybrid Neutrino Method}\label{implementation}
We have used the following parameters in a flat cosmology: $(\Omega_{\rm c} ~ + ~ \Omega_\nu,~ \Omega_{\rm b}, ~ \Omega_\Lambda,~ h,~\sigma_8, ~n_s) = (0.25,~ 0.05,~ 0.7,~ 0.7,~ 0.878,~ 1)$. The value for $\sigma_8$ is for a pure $\Lambda$CDM model without massive neutrinos. We use values of $\sum m_\nu = 0.6$ eV and 1.2 eV with particular emphasis on the latter value, since a high neutrino mass affect structure formation more, thereby testing the accuracy of the hybrid method better.

The primordial density perturbations are followed with the linearised Einstein and Boltzmann equations, solved in CAMB \cite{CAMB}. At $z=49$ the CDM component is taken out of CAMB and followed in \textsc{gadget}-2 \cite{Springel2} run in the TreePM mode. The $N$-body initial conditions are generated with the Zel'dovich Approximation \cite{Zeldovich:1969sb} as well as a second-order correction term \cite{Scoccimarro1} added to the CDM particles only.

\begin{figure}
   \noindent
   \begin{center}
   \hspace*{-1.0cm}\includegraphics[width=1.1\linewidth]{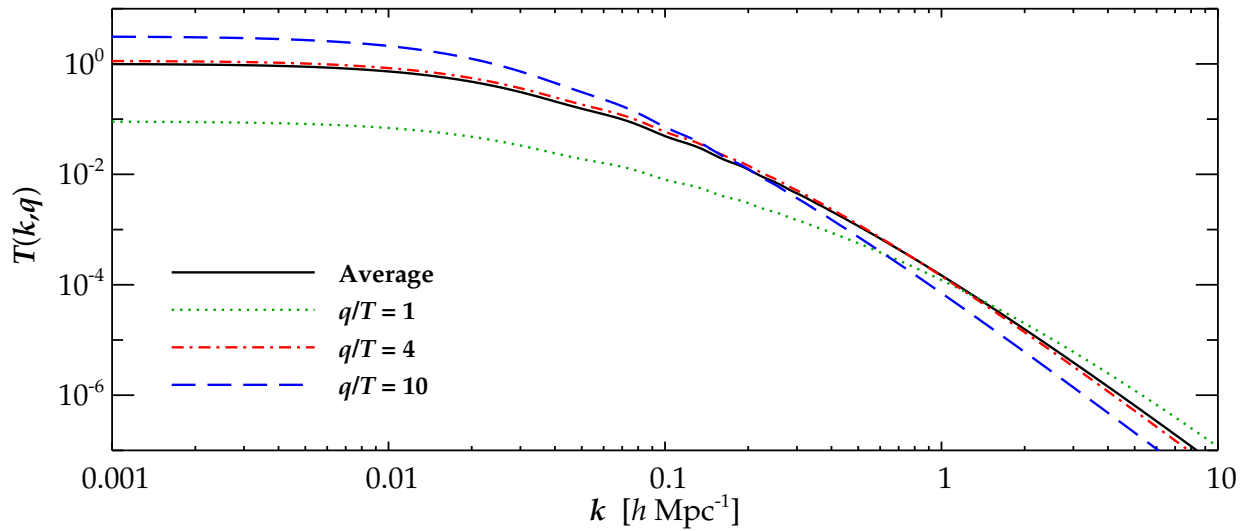}
   \end{center}
   \caption{Momentum dependent neutrino TFs at $z=0$ for $\sum m_\nu = 1.2 \, {\rm eV}$.}
   \label{fig:transfer}
\end{figure}

As input to the $N$-body code the linear neutrino TF is needed. Normally this TF is found from Eq.~(\ref{eq:drho}) by integrating over all momenta. However, in our case we need the momentum dependent TFs because when particles are extracted from the grid they are taken in a certain momentum range and the neutrino TFs are highly momentum dependent (because the free-streaming length depends on $q$). In Fig.~\ref{fig:transfer} we show the momentum dependent TFs for $\sum m_\nu = 1.2 \, {\rm eV}$ at $z = 0$, from which the neutrino momentum dependence on the power spectrum is clearly seen. The effect of free-streaming at small scales is most pronounced for the highest neutrino momentum states, but these states do have the largest TF at large scales.

To follow the neutrino component we combine the neutrino particle and grid representations, first explored in \cite{Brandbyge1} and \cite{Brandbyge2}, respectively. Initially, the neutrinos are represented on a grid, and there are no neutrino $N$-body particles. When the neutrino thermal velocity falls below a few times the average gravitational flow velocity, calculated from the CDM $N$-body particles, the lowest momentum part of the neutrino grid energy is converted into $N$-body particles. Here there are 3 things to consider.

First, the energy on the neutrino grid should be decreased. We have 15 neutrino TF bins (the standard output from CAMB, the number can easily be modified), where $q/T$ runs from 1 to 15 (upper limits) in integer steps, corresponding to a bin width of $\simeq 125 \, {\rm km / s}$ in comoving coordinates for $\sum m_\nu = 1.2 \, {\rm eV}$. When the first bin with the lowest neutrino momentum states is converted to particles, only the remaining 14 bins are summed to get the neutrino grid Fourier modes. The phases of the neutrino particles are made with the same set of random numbers as were the CDM particles at $z = 49$.

Second, the neutrino $N$-body particle mass should correspond to the mass removed from the grid.

Third, when neutrino $N$-body particles are created in a certain momentum range, the thermal velocity they recieve lies within this range as well. This is based on the fact that at the conversion redshift the thermal velocity is much larger than the gravitational flow velocity, meaning that neutrinos in a given bin can at most have originated from the adjacent bins. It is therefore a very good approximation to give the neutrino particles a thermal velocity corresponding to the bin and assuming that there is no correlation between the thermal and flow components. This reduced bin mixing caused by the thermal velocity is fortunate, since disentangling these two components would be impossible.

The thermal velocities for a given particle grid position in two succeeding neutrino grid-to-particle conversions are drawn with the same random numbers, but with momenta pointing in opposite directions. This ensures an approximate local conservation of momentum. In \cite{Brandbyge1} local conservation was not enforced due to a higher neutrino particle starting redshift, and since only the average TF was used as input power spectrum.

The neutrino gravitational velocity is found from the difference of two position grids centered around (with $\Delta a = 0.01$) our neutrino grid-to-particle conversion redshift (first described in \cite{Brandbyge1}). When we add the gravitational flow velocity, some neutrinos will have a total momentum which lies outside of the bin. This is a problem when part of the neutrino grid is converted to particles at times when the neutrino thermal velocity is only a few times the gravitational velocity, though effectively being a problem only for the low momentum neutrinos.

\begin{center}
\begin{table*}[t]
{\footnotesize %\small
  \hspace*{-0.0cm}\begin{tabular}
  {|c||c@{\hspace{2pt}}c@{\hspace{2pt}}c|c@{\hspace{2pt}}c@{\hspace{2pt}}c@{\hspace{2pt}}c@{\hspace{2pt}}c@{\hspace{2pt}}c|c@{\hspace{2pt}}c@{\hspace{2pt}}c@{\hspace{2pt}}c|c@{\hspace{2pt}}c@{\hspace{2pt}}c|}                      \hline
%%%%%%%%%%%%%%%%%
  & $A_1$ & $A_2$ & $A_3$ & $B_1$ & $B_2$ & $B_3$ & $B_4$ & $B_5$ & $B_6$ & $C_1$ & $C_2$ & $C_3$ & $C_4$ & $D_1$ & $D_2$ & $D_3$\\       \hline\hline
%%%%%%%%%%%%%%%%%
$N_{\rm CDM}$ & $512^3$ & $512^3$ & $512^3$ & $256^3$ & $256^3$ & $256^3$ & $256^3$ & $256^3$ & $256^3$ & $256^3$ & $256^3$ & $256^3$ & $256^3$&$256^3$&$256^3$&$256^3$\\
%%%%%%%%%%%%%%%%%
$N_{\nu,{\rm grid}}$ &$512^3$ & $0$& $512^3$ & $256^3$ & $256^3$ & $256^3$&$256^3$&$256^3$&$256^3$&$256^3$ & $256^3$&$256^3$& $256^3$& $256^3$ & $256^3$ & $256^3$\\
%%%%%%%%%%%%%%%%%
$q_{\rm cut}/T$ &10    & $\infty$ & $0$& 4 & $5$&6&7&$10$&$15$&$10$   & $10$& $10$ & $10$  & 10 &10 & $10$\\
%%%%%%%%%%%%%%%%%
$f_{\rm flow}$       & 4& $\infty$&0& 4&4&4&4&4&4&2&4&8&16 &2& 4&8\\
%%%%%%%%%%%%%%%%%
$\sum m_\nu~[{\rm eV}]$ & 1.2& 1.2&1.2&1.2&1.2&1.2&1.2&1.2&1.2&1.2&1.2&1.2&1.2&0.6&0.6 &0.6  \\
%%%%%%%%%%%%%%%%%
$\Omega_{\nu}$~[$\%$] & 2.6& 2.6&2.6& 2.6&2.6&2.6&2.6&2.6&2.6&2.6&2.6&2.6&2.6&1.3 &1.3& 1.3\\\hline
  \end{tabular}
  }
  \caption{Parameters for our $N$-body simulations presented in this paper. $N_{\rm CDM}$ is the number of CDM $N$-body particles and $N_{\nu,{\rm grid}}$ is the size of the neutrino Fourier grid. $q_{cut}/T$ indicates how much of neutrino momentum space is converted to particles, and the number of neutrino $N$-body particles can be found from $q_{cut}/T \cdot N_{\nu,{\rm grid}}$. When the average CDM gravitational flow velocity times $f_{\rm flow}$ has increased above $q/T = 1$, the conversion from grid to particles is made. $\sum m_\nu $ is the total neutrino mass, and it is in all cases related to the one-particle neutrino mass, $m_\nu$, by $\sum m_\nu = 3 m_\nu$. $\Omega_{\nu}$ is the fraction of the critical density contributed by the neutrinos today. In all simulations $R_{\rm BOX} = 512 h^{-1} {\rm Mpc}$, and the size of the particle mesh grid is equal to the number of CDM particles. All simulations have a starting redshift of 49.}
  \label{fig:table1}
\end{table*}
\end{center}

\section{Results}\label{results}
%%%%%%%%%%%%%%%%%%%%%%%%%%% figures %%%%%%%%%%%%%%%%%%%%%%%%%%
\begin{figure}
   \noindent
     \begin{center}
     \includegraphics[width=0.65\linewidth]{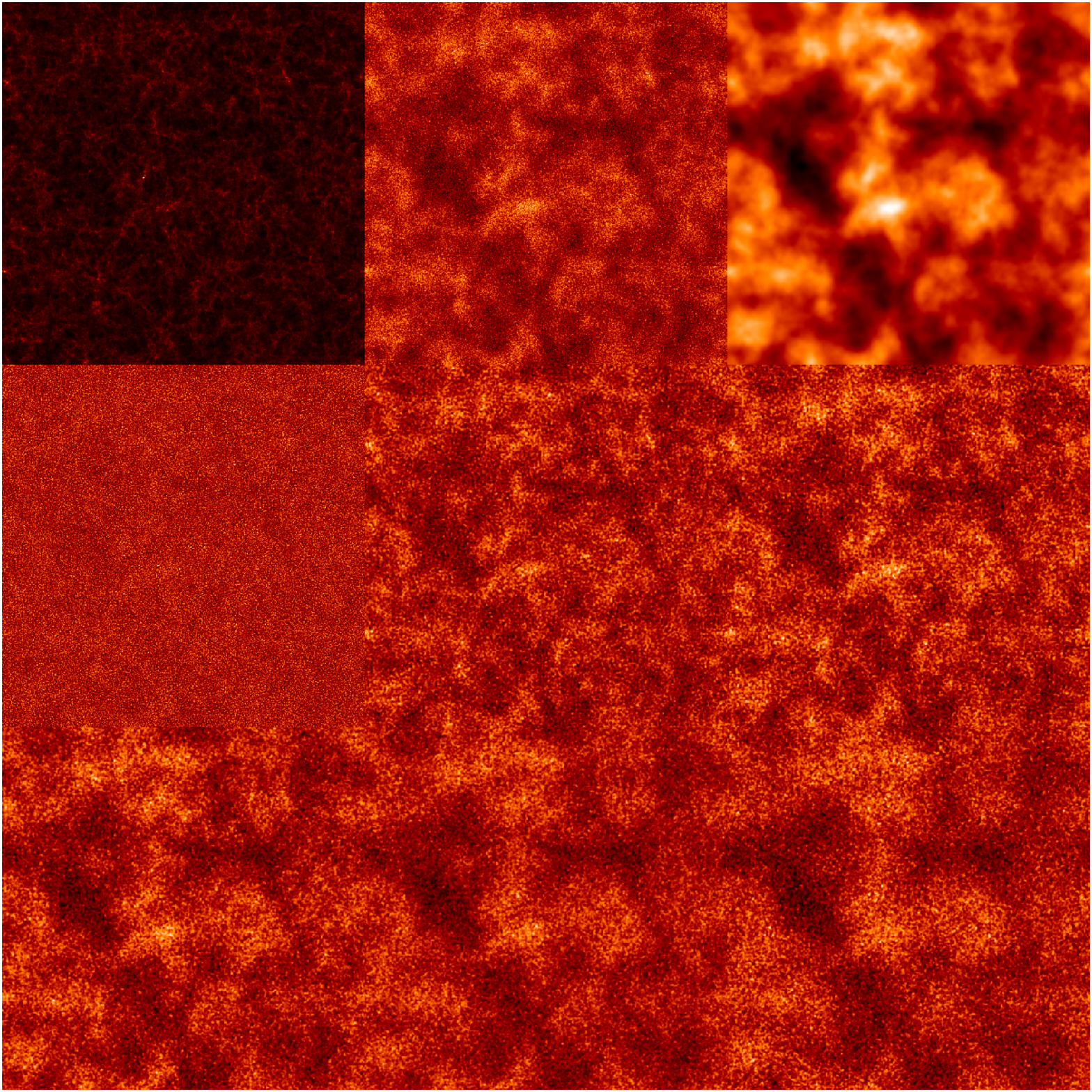}\\
     \end{center}
     \vfill
     \begin{center}
     \includegraphics[width=0.65\linewidth]{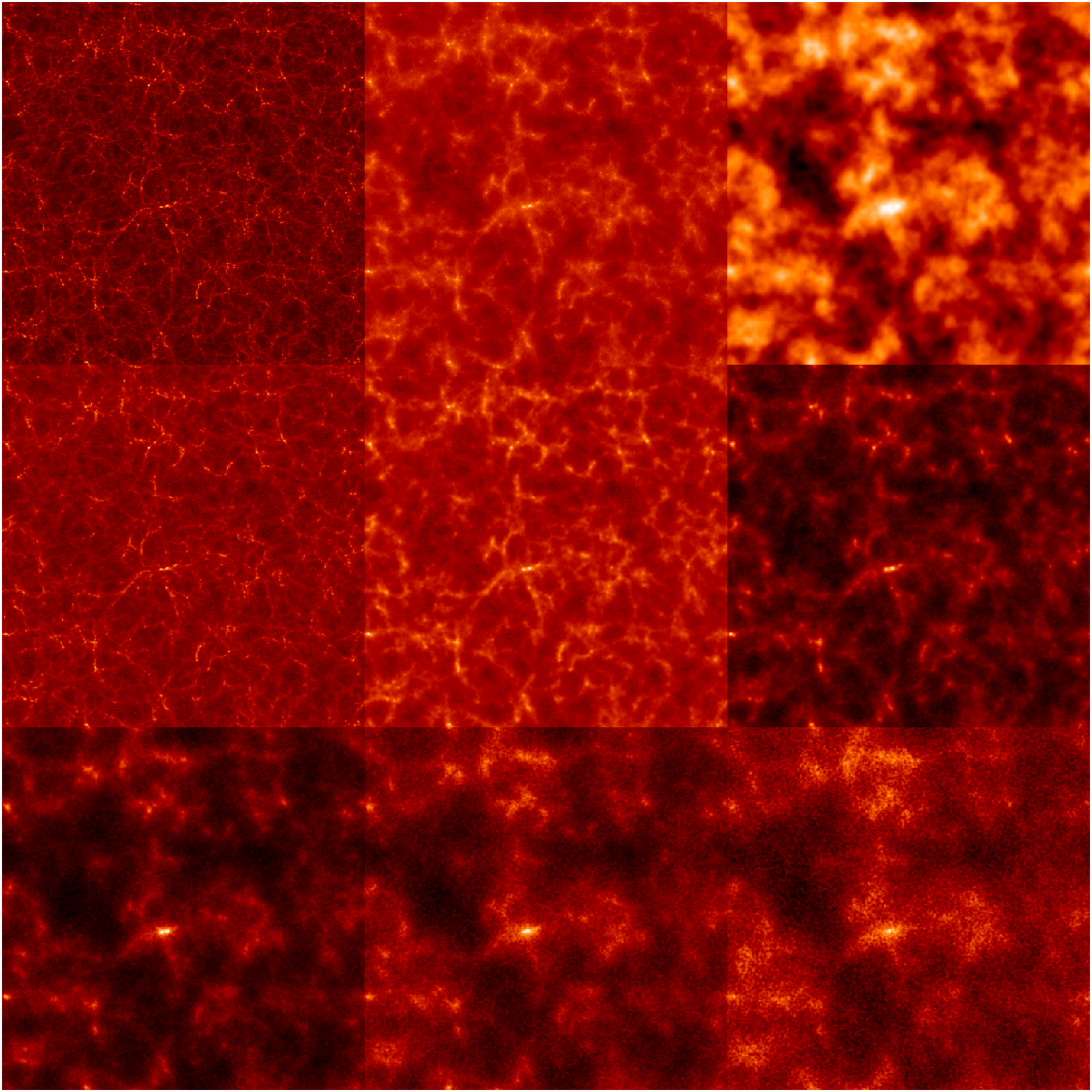}
     \end{center}
   \caption{Density grids from the hybrid simulation $A_1$ with $\sum m_\nu = 1.2 \, {\rm eV}$ found with the adaptive smoothing length kernel from \cite{monaghan}. Top mosaic: $z = 4$. Bottom mosaic: $z = 0$. Top row: CDM, neutrino particles and neutrino grid. Middle row: $q/T = $ 1, 2 and 3. Bottom row: $q/T = $ 5, 7 and 10. Bottom mosaic: The CDM, neutrino particle, $q/T = $ 1, 2 and 3 density grids have been raised to the power of $0.25$ to enhance the dynamical contrast. The density slices have a thickness of $20 h^{-1} \, {\rm Mpc}$ and are $512 h^{-1} \, {\rm Mpc}$ on a side.}
   \label{fig:rho}
\end{figure}

In Fig.~\ref{fig:rho} we show density grids from the hybrid $A_1$ simulation (see Table~\ref{fig:table1}). Some features can immediately be seen from this. As expected the part of the neutrino distribution remaining on the grid is much less clustered on small scales than the neutrino particles, while the opposite is true on the largest scales. Interestingly, it can also be seen that the $q/T=1$ bin is initially only weakly clustered on all simulated scales (i.e.\ at $z=4$), but quickly starts tracking the CDM component, and at $z=0$ has as much clustering as the next momentum bin. Note that the individual neutrino density grids are found from the original binning of the neutrino particles at the grid-to-particle conversion redshift (determined by the primordial thermal velocity), and not according to the actual neutrino particle velocities at redshifts 4 and 0.

\subsection{Comparing the hybrid method to its building blocks}
We have compared our new hybrid method ($A_1$) with the full non-linear simulation ($A_2$), where neutrino $N$-body particles are created at $z = 49$, and with the simulation where the neutrino component stays linear on a grid ($A_3$).

\begin{figure}
   \noindent
   \begin{minipage}{0.49\linewidth}
     \hspace*{-0.5cm}\includegraphics[width=1.1\linewidth]{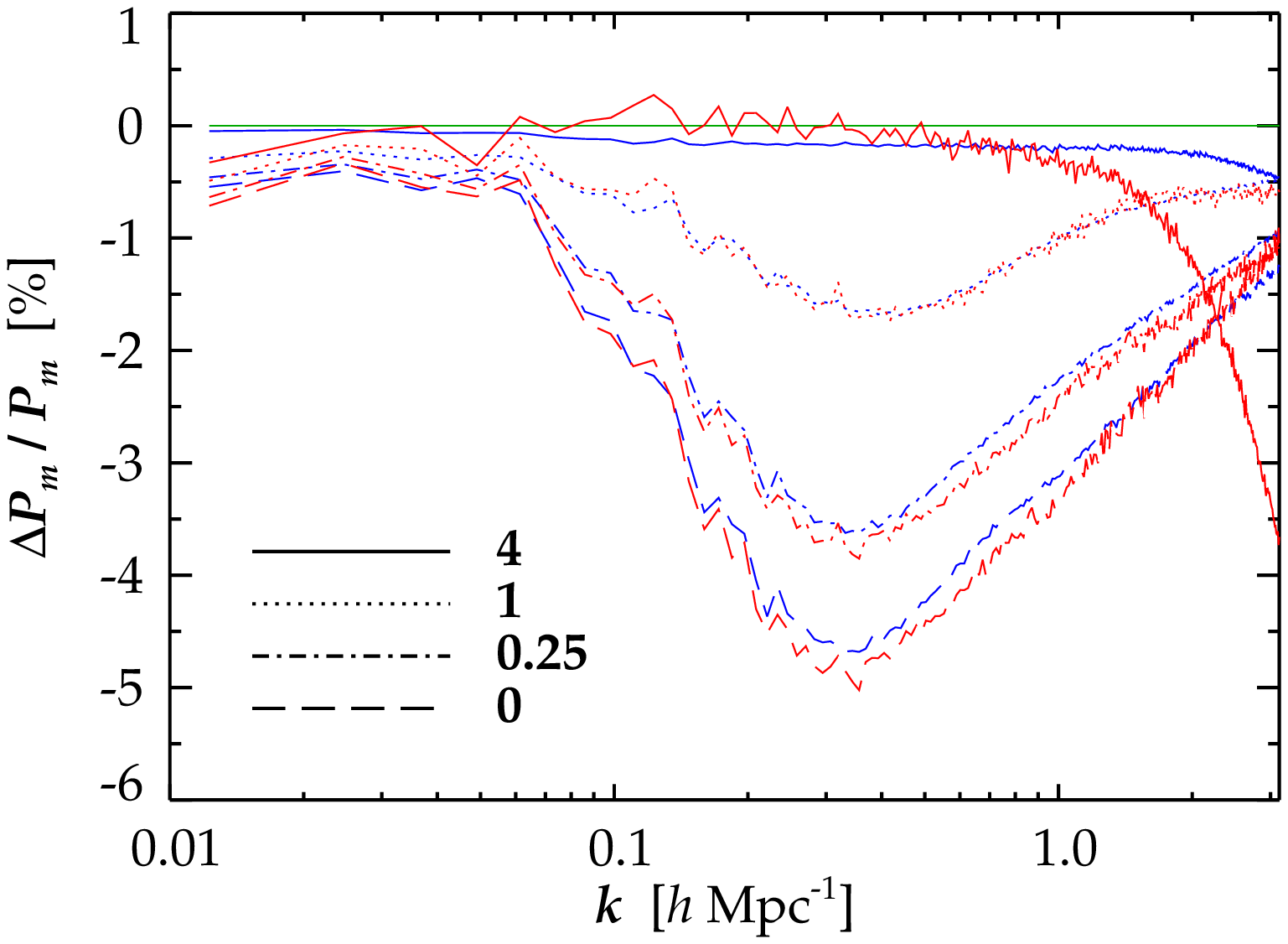}
   \end{minipage}
   \begin{minipage}{0.49\linewidth}
     \hspace*{-0.0cm}\includegraphics[width=1.1\linewidth]{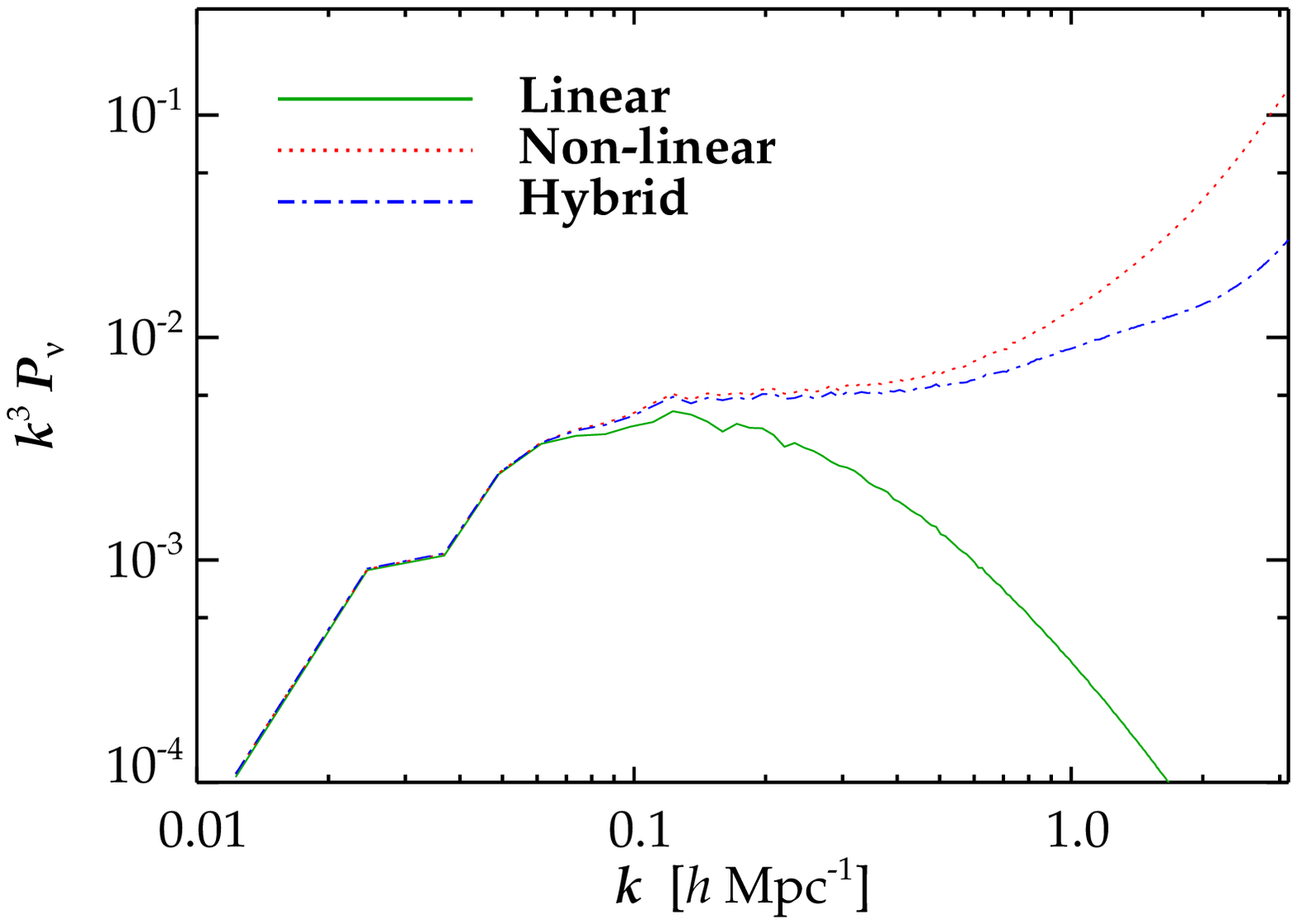}
   \end{minipage}
   \caption{Left: Difference between the linear simulation $A_3$ (green) and the full non-linear $A_2$ (red) and hybrid $A_1$ (blue) methods at various redshifts. Right: Absolute neutrino power spectra at $z = 0$.}
   \label{fig:power0_8}
\end{figure}

On the left hand side of Fig.~\ref{fig:power0_8}, it can be seen that the new method gives almost the same total matter power spectrum as the full non-linear simulation, except at low redshift where the hybrid method gives slightly less power for $k \gtrsim 0.2 \, h {\rm Mpc}^{-1}$. In the full non-linear simulation it can be seen from \cite{Brandbyge2} that the matter power spectrum has not converged for $512^3$ neutrino particles. The discrepancy seen in Fig.~\ref{fig:power0_8} is therefore due to a complete suppression of the noise term in the hybrid simulation, which we have verified by running simulations with $10\cdot 256^3$ ($C_2$) and $10\cdot 512^3$ ($A_1$) neutrino $N$-body particles with the hybrid method.

On the right hand side of Fig.~\ref{fig:power0_8} we show the neutrino power spectrum for the different methods. With the hybrid method we are now able to simulate the non-linear neutrino power spectrum accurately out to $k \simeq 1 \, h {\rm Mpc}^{-1}$, even with $R_{\rm BOX} = 512 \, h^{-1} \, {\rm Mpc}$. The extra accuracy is of course achieved by including more neutrino particles, but with a comparable amount of CPU time.

\subsection{Converting part of neutrino momentum space to particles}
Since only a small fraction of the neutrino mass resides in the high momentum tail it is desirable to find a momentum cut-off above which the neutrinos do not contribute significantly to structure formation at a given scale. Such a cut-off will reduce the simulation time, because the timestep criterion depends in part on the total velocity of the particles, and particles in the highest momentum bins have thermal velocities much higher than the gravitional flow velocities even at $z = 0$.

\begin{figure}
   \noindent
   \begin{minipage}{0.49\linewidth}
     \hspace*{-0.5cm}\includegraphics[width=1.1\linewidth]{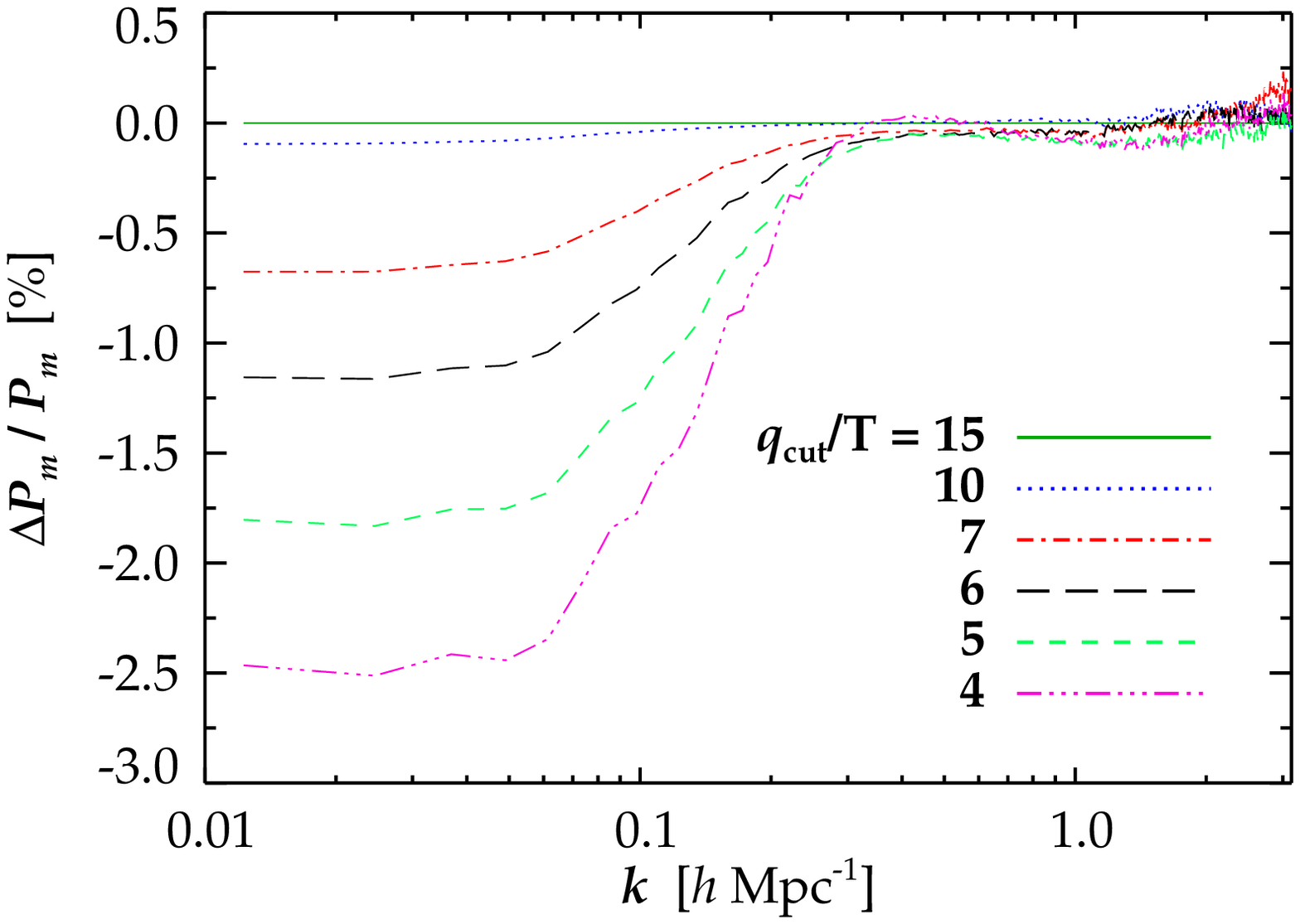}
   \end{minipage}
   \begin{minipage}{0.49\linewidth}
     \hspace*{-0.0cm}\includegraphics[width=1.1\linewidth]{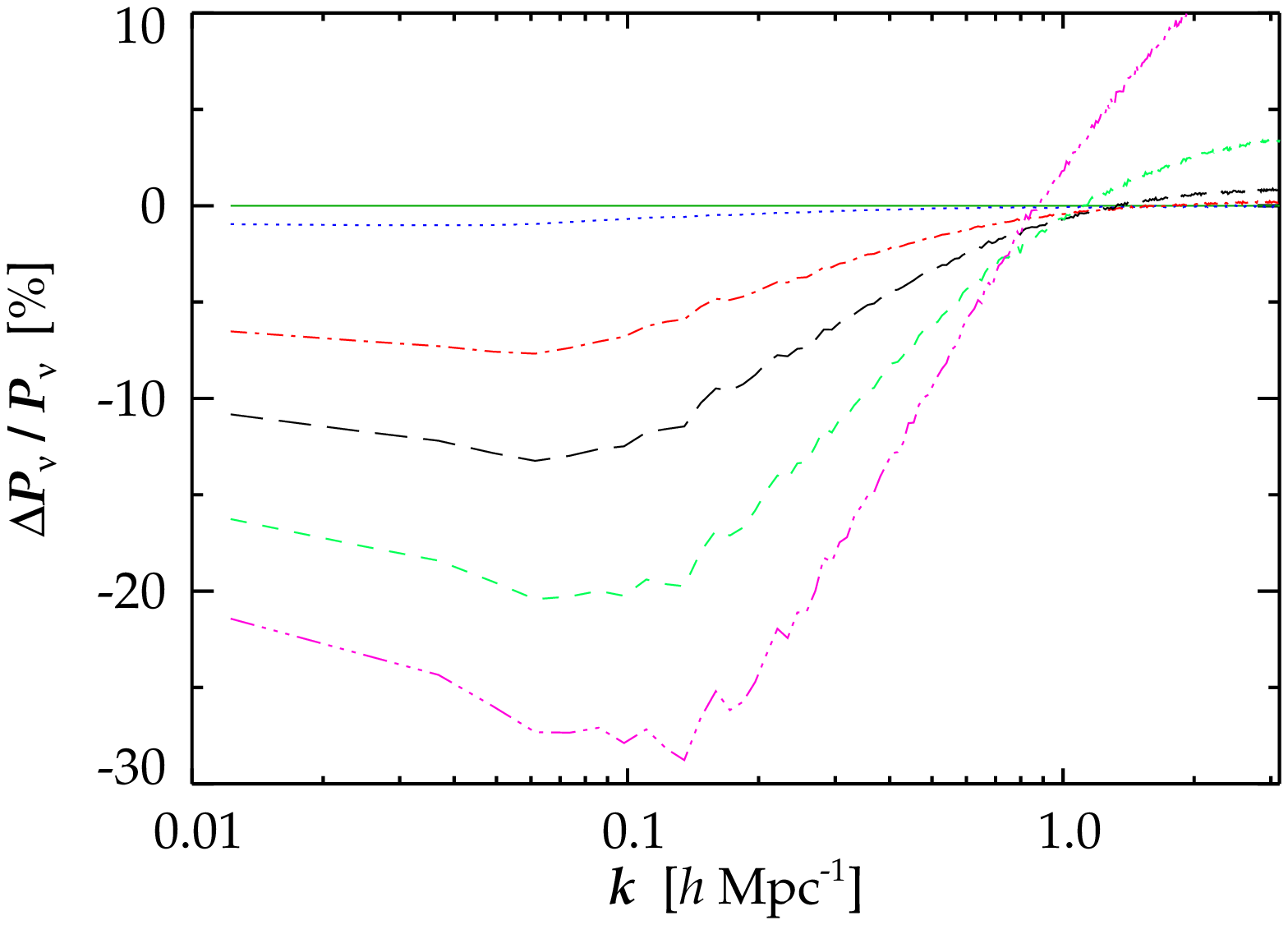}
   \end{minipage}
   \caption{$z = 0$. Left: Hybrid simulations $B_i$ with $q_{\rm cut}/T$ ranging from 4 to 15 compared to $q_{\rm cut}/T = 15$ ($B_6$). Right: Corresponding neutrino power spectra.}
   \label{fig:power1_2}
\end{figure}

In Fig.~(\ref{fig:power1_2}) we show the total matter (left) and neutrino (right) power spectra for simulations with a cut-off at $q_{\rm cut}/T = 4$, 5, 6, 7, 10 and 15. For $q_{\rm cut}/T = 5$ and 10, 86\% and 99.7\% of the neutrino mass is converted to particles, respectively. The total matter power spectrum is almost identical for $q_{\rm cut}/T=10$ and 15. For $q_{\rm cut}/T = 5$ the matter power spectrum increases by 2\%. This is due to leakage of power across the momentum cut-off scale. But a 1\% error in the large-scale perturbation field hardly affect the formation of small-scale structure. For cases (such as the study of halo profiles) where only the small-scale structure of neutrinos is important, $q_{\rm cut}/T \simeq 5$ would be sufficient.

In the neutrino power spectra the same trends can be identified. There is no significant double counting of perturbations at small scales, since at these scales the linear grid does not contribute for $q > q_{\rm cut}$. The fact that the $q_{\rm cut}/T \lesssim 6$ simulations give less power at $k \gtrsim 1 \, h {\rm Mpc}^{-1}$, can be attributed to both neutrino $N$-body particle shot noise and a lack of non-linear corrections from the non-converted grid bins. We leave the relative importance of these two effects to people with too much CPU time.

\subsection{The optimal grid-to-particle conversion redshift}
The optimal grid-to-particle conversion redshift is found as a trade-off between ensuring that non-linearities in the neutrino component is accurately simulated and that the finite number of neutrino $N$-body particles do not introduce a significant noise term. The first requirement is satisfied by a high and the second by a low conversion redshift.

\begin{figure}
   \noindent
   \begin{minipage}{0.49\linewidth}
     \hspace*{-0.5cm}\includegraphics[width=1.1\linewidth]{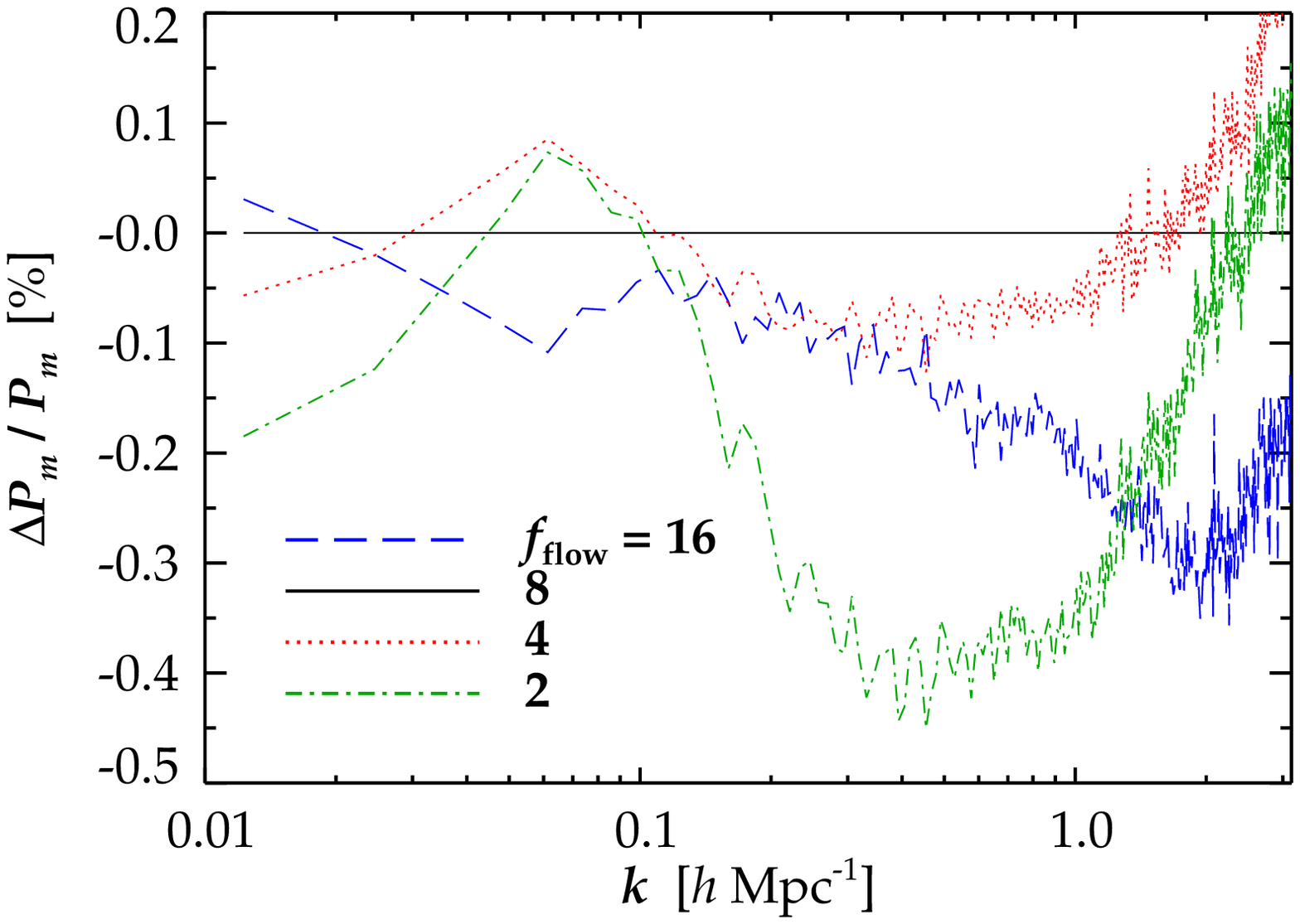}
   \end{minipage}
   \begin{minipage}{0.49\linewidth}
     \hspace*{-0.0cm}\includegraphics[width=1.1\linewidth]{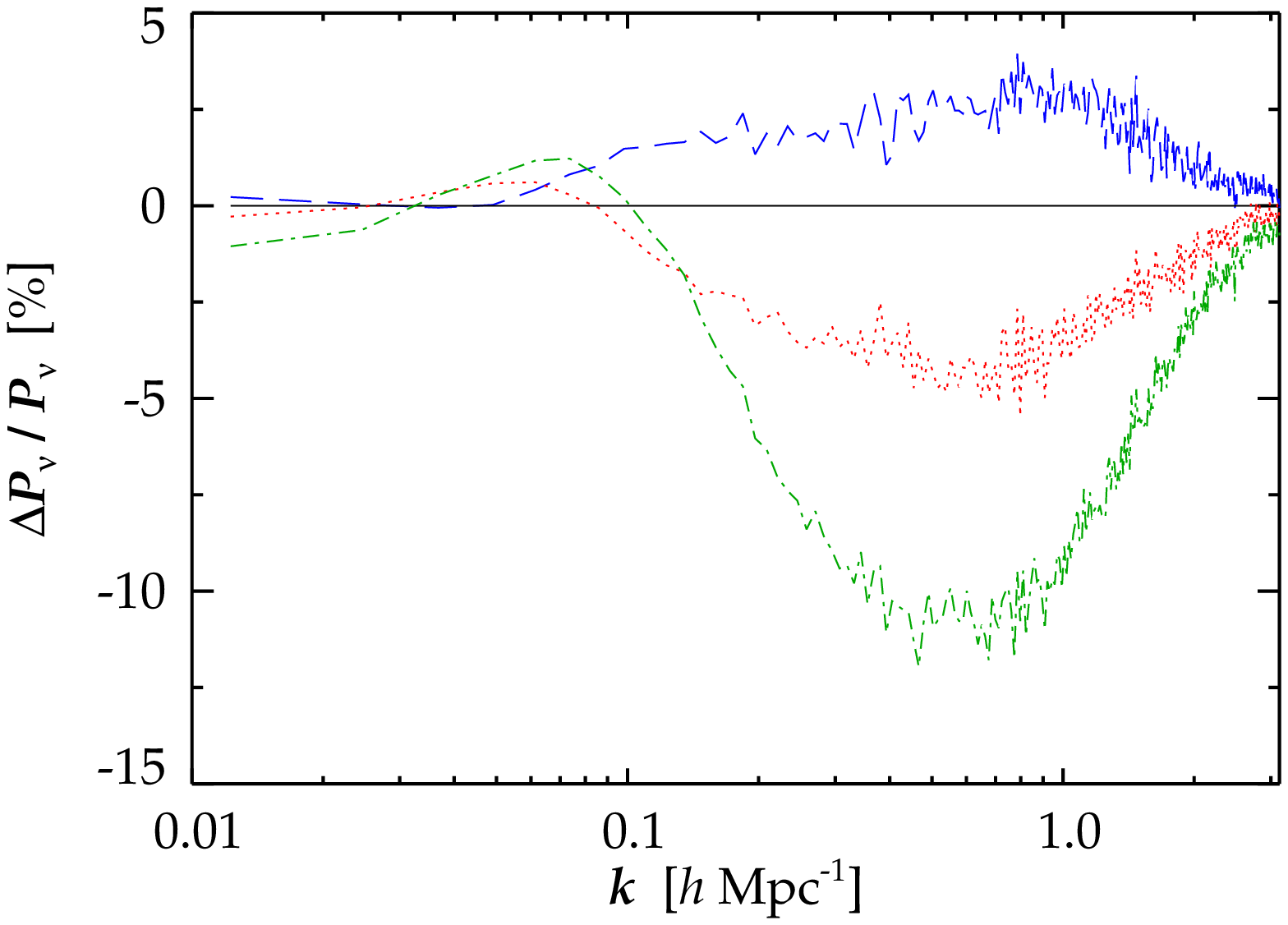}
   \end{minipage}
   \caption{$\sum m_\nu = 1.2 \, {\rm eV}$. Left: Matter power spectra from simulations $C_i$ at $z = 0$ with grid-to-particle conversion factors of $f_{\rm flow} = 2$, 4, 8 and 16 relative to $f_{\rm flow} = 8$ ($C_3$). Right: Corresponding neutrino power spectra.}
   \label{fig:power3_4}
\end{figure}

\begin{figure}
   \noindent
   \begin{minipage}{0.49\linewidth}
     \hspace*{-0.5cm}\includegraphics[width=1.1\linewidth]{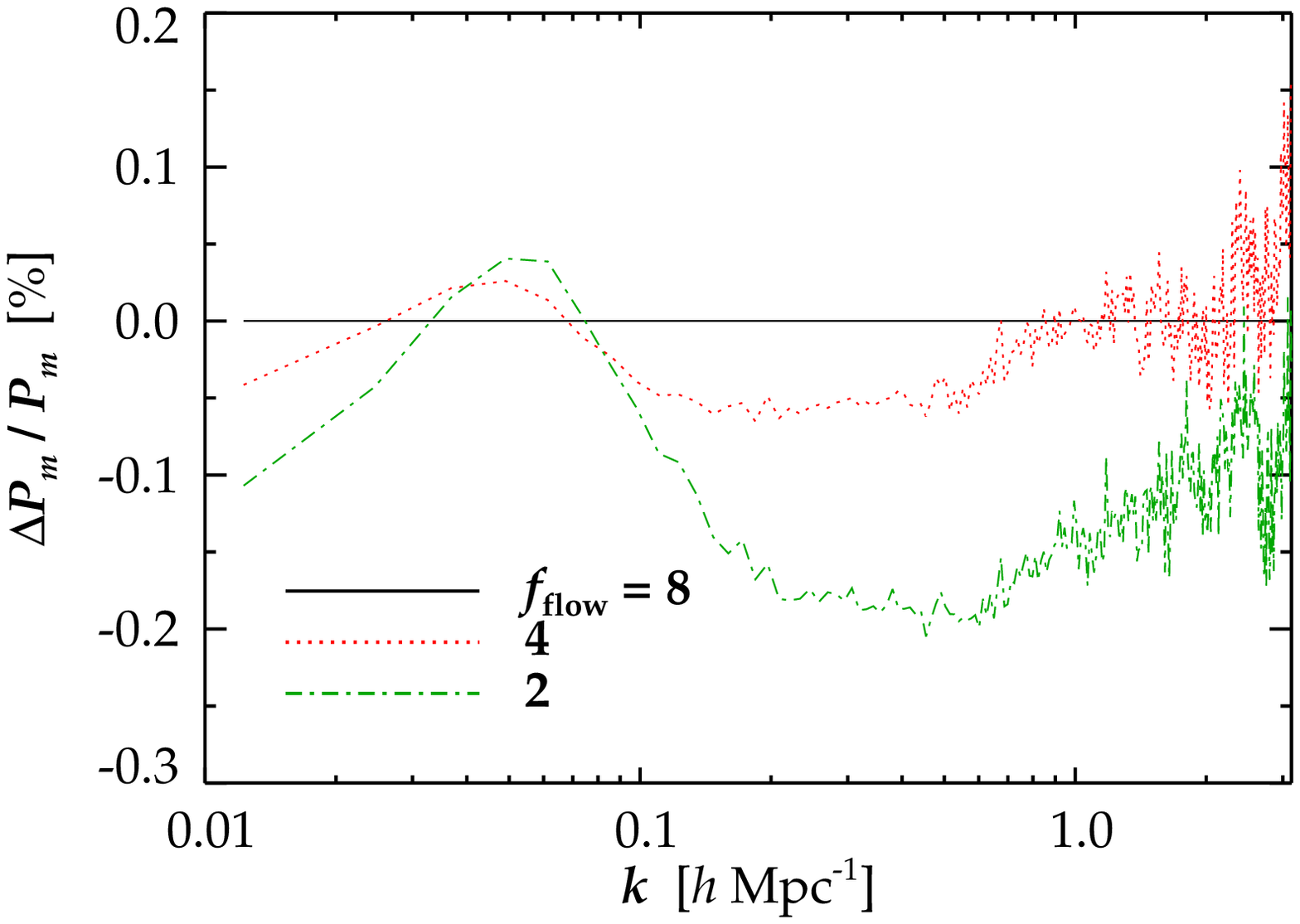}
   \end{minipage}
   \begin{minipage}{0.49\linewidth}
     \hspace*{-0.0cm}\includegraphics[width=1.1\linewidth]{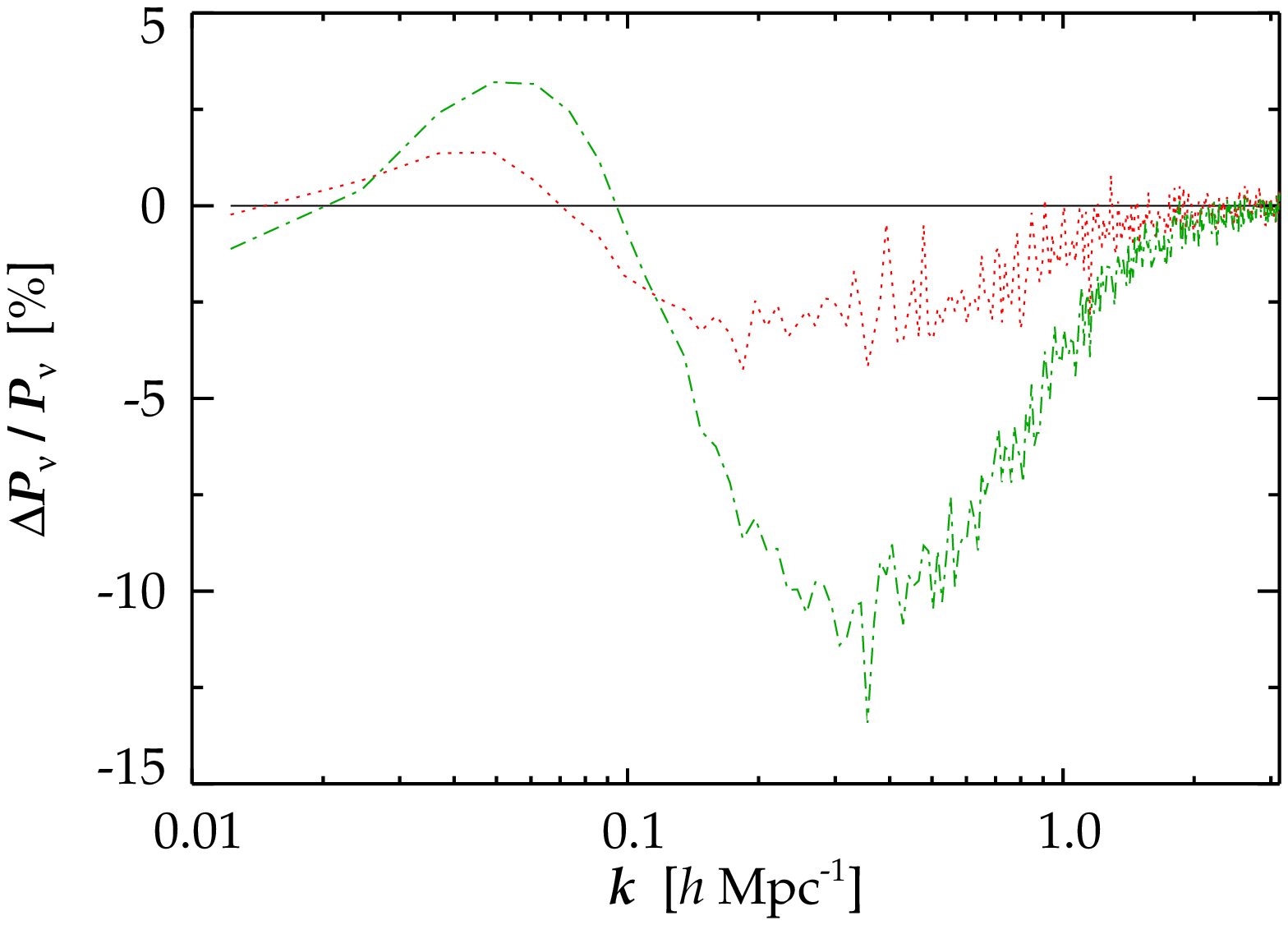}
   \end{minipage}
   \caption{$\sum m_\nu = 0.6 \, {\rm eV}$. Left: Matter power spectra from simulations $D_i$ at $z = 0$ with grid-to-particle conversion factors of $f_{\rm flow} = 2$, 4 and 8 relative to $f_{\rm flow} = 8$ ($D_3$). Right: Corresponding neutrino power spectra.}
   \label{fig:power6_7}
\end{figure}

In Fig.~(\ref{fig:power3_4}) we show the difference in the matter and neutrino power spectra with $\sum m_\nu = 1.2 \, {\rm eV}$ for different conversion redshifts. The conversion criterion used was when the upper velocity limit of the $q/T=1$ momentum bin had fallen below some factor $f_{\rm flow}$ of the average CDM $N$-body particle flow velocity. This quantity is related to the strength of the gravitational field, and therefore to the ability of neutrino particles to cluster beyond linear order. We have run simulations with $f_{\rm flow} = 2$, 4, 8 and 16, corresponding to conversion redshifts of 3.1, 5.5, 9.4 and 15.7, respectively. It can be seen that with $f_{\rm flow} = 8$ the matter power spectrum can be found with 0.2\% precision and the neutrino power spectrum with at precision at the 2-4\% level. The simulation with $f_{\rm flow} = 2$ clearly lacks non-linear corrections due to the late conversion redshift.

By running simulations with the particle grids down-scaled by $2^3$ we have found that these results have not yet converged, though they have converged at an acceptable level to demonstrate the robustness of the hybrid method. Likewise, we have found that the lower value of the matter power spectrum with $f_{\rm flow} = 16$ compared to $f_{\rm flow} = 8$ is a noise term, making the neutrino distributions decorrelate with its CDM counterpart, giving rise to less structure.

In Fig.~(\ref{fig:power6_7}) the same case is examined with $\sum m_\nu = 0.6 \, {\rm eV}$, and the same general trends can be identified. The relative error in the matter power spectrum is proportional to $\Omega_\nu$ and the relative error in the neutrino power spectrum is roughly independent of $\Omega_\nu$.

\section{Discussion and Conclusions}\label{conclusions}

We have presented a new scheme for resolving the small-scale structure of neutrinos. It combines features from grid simulations \cite{Brandbyge1} which are fast and accurately track neutrinos in the linear regime with particle simulations \cite{Brandbyge2} capable of resolving neutrino structures in the highly non-linear regime.

The hybrid scheme is ideal for calculating the non-linear matter power spectrum for neutrino masses around the current upper bound $\sum m_\nu \simeq 0.5 \, {\rm eV}$. For lower neutrino masses the grid approach has an accuracy better than 1\%. Furthermore, the hybrid scheme with separate momentum dependent initial TFs and corresponding separate thermal velocities reproduces our earlier results from \cite{Brandbyge1} and \cite{Brandbyge2}, where the average neutrino TF was used.

For $\sum m_\nu = 1.2 \, {\rm eV}$ converting 5 to 10 momentum bins at redshifts $z \simeq 5 - 10$ will accurately reproduce the matter power spectrum together with the neutrino power spectrum on small scales. For lower neutrino masses the particles can be created later due to their higher thermal velocities and a smaller part of the neutrino energy needs to be converted to particles to accurately simulate neutrino small-scale structure.

In the present work we have focused on features in the power spectrum and limited our discussion to scales of $k \lesssim 1 \, h \, {\rm Mpc}^{-1}$, where our code can easily keep errors under control at the 1\% level. Achieving a comparable level of accuracy in pure particle-based simulations would require a prohibitive amount of CPU time.

Even more interestingly, this hybrid scheme is capable of resolving much smaller neutrino structures, such as galaxy- or cluster-sized halos. This will be the topic of a follow-up paper \cite{bhhw09}.

%%%%%%%%%%%%%%%%%%%%%%%%%%%%%%%%%%%%%%%%%%%%%%%%%%%%%%%%%%%%%%%%%%%%%%%%%%%%%%%%%%%%%%%%%%%%%%%%%%%%%%%%%%%%%%%%%%%%%%%%%%%%%%%%%%%%%%
%%%%%%%%%%%%%%%%%%%%%%%%%%%%%%%%%%%%%%%%%%%%%%%%%%%%%%%%%%%%%%%%%%%%%%%%%%%%%%%%%%%%%%%%%%%%%%%%%%%%%%%%%%%%%%%%%%%%%%%%%%%%%%%%%%%%%%
\section*{Acknowledgements}
We acknowledge computing resources from the Danish Center for
Scientific Computing (DCSC). We thank Troels Haugb{\o}lle and Yvonne Wong for discussions and comments on the manuscript.

%%%%%%%%%%%%%%%%%%%%%%%%%%%%%%%%%%%%%%%%%%%%%%%%%%%%%%%%%%%%%%%%%%%%%%%%%%%%%%%%%%%%%%%%%%%%%%%%%%%%%%%%%%%%%%%%%%%%%%%%%%%%%%
%%%%%%%%%%%%%%%%%%%%%%%%%%%%%%%%%%%%%%%%%%%%%%%%%%%%%%%%%%%%%%%%%%%%%%
%%%%%%%%%%%%%%%%%%%%%%%%%%%%%%%%%%%%%%%%%%%%
\section*{References} %%%%%%%%%%%%%%%%%%%%%%%%%%%%%%%%%%%%%%%%%%%%%%%%
%%%%%%%%%%%%%%%%%%%%%%%%%%%%%%%%%%%%%%%%%%%%%%%%%%%%%%%%%%%%%%%%%%%%%%

%%%%%%%%%%%%%%%%%%%%%%%%%%%%%%%%%%%%%%%%%%%%%%%

\begin{thebibliography}{00}
%%%%%%%%%%%%%%%%%%%%%%%%%%%%%%%%%%%%%%%%%%%%

%\cite{Schwetz:2008er}
\bibitem{Schwetz:2008er}
  T.~Schwetz, M.~Tortola and J.~W.~F.~Valle,
  %``Three-flavour neutrino oscillation update,''
  New J.\ Phys.\  {\bf 10}, 113011 (2008)
  [arXiv:0808.2016 [hep-ph]].
  %%CITATION = NJOPF,10,113011;%%

%
\bibitem{Wong:2008ws}
  Y.~Y.~Y.~Wong,
  %``Higher order corrections to the large scale matter power spectrum in the
  %presence of massive neutrinos,''
  JCAP {\bf 0810}, 035 (2008)
  [arXiv:0809.0693 [astro-ph]].
  %%CITATION = JCAPA,0810,035;%%

%\cite{Lesgourgues:2009am}
\bibitem{Lesgourgues:2009am}
  J.~Lesgourgues, S.~Matarrese, M.~Pietroni and A.~Riotto,
  %``Non-linear Power Spectrum including Massive Neutrinos: the Time-RG Flow
  %Approach,''
  JCAP {\bf 0906}, 017 (2009)
  [arXiv:0901.4550 [astro-ph.CO]].
  %%CITATION = JCAPA,0906,017;%%

%\cite{Saito:2009ah}
\bibitem{Saito:2009ah}
  S.~Saito, M.~Takada and A.~Taruya,
  %``Nonlinear power spectrum in the presence of massive neutrinos: perturbation
  %theory approach, galaxy bias and parameter forecasts,''
  arXiv:0907.2922 [astro-ph.CO].
  %%CITATION = ARXIV:0907.2922;%%

\bibitem{Brandbyge1}
  J.~Brandbyge, S.~Hannestad, T.~Haugb{\o}lle and B.~Thomsen,
  JCAP {\bf 0808} (2008) 020
  [arXiv:0802.3700 [astro-ph]].

\bibitem{Brandbyge2}
  J.~Brandbyge, S.~Hannestad,
  JCAP {\bf 0905} (2009) 002
  [arXiv:0812.3149 [astro-ph]].

%\cite{Ma}
\bibitem{Ma}
  C.~P.~Ma and E.~Bertschinger,
  %``Cosmological perturbation theory in the synchronous and conformal Newtonian
  %gauges,''
  Astrophys.\ J.\  {\bf 455}, 7 (1995)
  [arXiv:astro-ph/9506072].
  %%CITATION = ASJOA,455,7;%%

%\cite{Lewis:2002ah}
\bibitem{CAMB}
  A.~Lewis and S.~Bridle,
  Phys.\ Rev.\ D {\bf 66}, 103511 (2002)
  [arXiv:astro-ph/0205436].
  %%CITATION = ASTRO-PH 0205436;%%

%\cite{Springel:2005mi}
\bibitem{Springel2}
  V.~Springel,
  %``The cosmological simulation code GADGET-2,''
  Mon.\ Not.\ Roy.\ Astron.\ Soc.\  {\bf 364}, 1105 (2005)
  [arXiv:astro-ph/0505010].
  %%CITATION = MNRAA,364,1105;%%

%\cite{Zeldovich:1969sb}
\bibitem{Zeldovich:1969sb}
  Y.~B.~Zeldovich,
  %``Gravitational instability: An Approximate theory for large density
  %perturbations,''
  Astron.\ Astrophys.\  {\bf 5}, 84 (1970).
  %%CITATION = AAEJA,5,84;%%

%\cite{Scoccimarro:1997gr}
\bibitem{Scoccimarro1}
  R.~Scoccimarro,
  %``Transients from Initial Conditions: A Perturbative Analysis,''
  Mon.\ Not.\ Roy.\ Astron.\ Soc.\  {\bf 299}, 1097 (1998)
  [arXiv:astro-ph/9711187].
  %%CITATION = MNRAA,299,1097;%%

\bibitem{monaghan}J.~J.~Monaghan and J.~C.~Lattanzio, Astron. Astrophys. {\bf 149}, 135 (1985).

\bibitem{bhhw09}
  J.~Brandbyge, S.~Hannestad, T.~Haugb{\o}lle, and Y.Y.Y.~Wong, in preparation.

%%%%%%%%%%%%%%%%%%%%%%%%%%%%%%%%%%%%%%%%%%%%
\end{thebibliography}
\end{document}